\documentclass[superscriptaddress,twocolumn,showpacs,amsmath,amssymb]{revtex4}

\usepackage[dvips]{graphicx}
\usepackage{graphicx}

\begin{document}

\title{
Role of non-collective excitations in 
low-energy heavy-ion reactions}

\author{S. Yusa}
\affiliation{
Department of Physics, Tohoku University, Sendai 980-8578,  Japan} 

\author{K. Hagino}
\affiliation{
Department of Physics, Tohoku University, Sendai 980-8578,  Japan} 

\author{N. Rowley}
\affiliation{
Institut de Physique Nucl\'{e}aire, UMR 8608, CNRS-IN2P3 et Universit\'{e}
de Paris Sud, 91406 Orsay Cedex, France}


\begin{abstract}
We investigate the effect of single-particle excitations on 
heavy-ion reactions 
at energies   
near the Coulomb barrier. 
To this end, we describe 
single-particle degrees of freedom with 
the random matrix theory and solve the 
coupled-channels equations 
for one-dimensional systems. 
We find that the single-particle excitations hinder the penetrability 
at energies above the barrier, leading to a smeared barrier 
distribution. 
This indicates that the single-particle excitations provide a 
promising way to explain the difference in a quasi-elastic 
barrier distribution recently observed in 
$^{20}$Ne + $^{90,92}$Zr systems. 
\end{abstract}


\maketitle

\section{Introduction}

Heavy-ion reactions near the Coulomb barrier provide a good
opportunity to investigate an interplay 
between the reaction process and 
internal excitations in the colliding nuclei. 
For example, 
it is well known that subbarrier fusion cross sections are
significantly enhanced 
as compared to a prediction of a simple barrier penetration model 
because of the couplings 
of the relative motion between the colliding nuclei to nuclear intrinsic 
degrees of freedom
\cite{dasgupta}.
It has been well recognized by now that 
the enhancement of fusion cross sections can be explained 
in terms of a 
distribution of Coulomb barrier heights originated from 
the couplings \cite{fusionbar}.
The barrier distribution can be actually extracted directly 
from the experimental data 
by taking 
the second derivative
of the product of fusion cross section $\sigma_{\rm fus}$ and the center of 
mass energy $E$ with respect to $E$, 
that is, $d^2\left(E\sigma_{\rm fus}(E)\right) / dE^2$ 
\cite{fusionbar}. 
The experimental data have clearly shown that the barrier distribution 
makes a useful representation to understand the reaction 
dynamics of heavy-ion subbarrier fusion reactions \cite{dasgupta,fusionexp}.

A similar concept of barrier distribution 
has been applied also to 
heavy-ion quasi-elastic scattering (that is, a sum of elastic, inelastic 
scattering, and transfer reactions) \cite{timmers,qelbar}.
In this case, 
the barrier distribution is defined as the first derivative of the 
ratio of the 
quasi-elastic scattering cross section at a backward angle 
$\sigma_{\rm qel}$ to the 
Rutherford cross section $\sigma_{\rm R}$, that is,  
$-d\left(\sigma_{\rm qel}/\sigma_{\rm R}\right)/dE$.
The fusion and quasi-elastic barrier distributions 
have been found to behave similarly to each other at least 
in a qualitative way \cite{timmers,zamrun}. 

In order to analyse the subbarrier enhancement of fusion 
cross sections and fusion and quasi-elastic barrier distributions, 
the coupled-channels method has been 
successfully employed. Typically,  
a few low-lying collective excitations are taken 
into account in a calculation\cite{dasgupta,ccfull}. 
With this approach, the barrier distribution arises 
naturally through the eigen-channel representation 
\cite{dasgupta,dasso2,HTB97,NBT86}. 

Recently, however, a few experimental data which cannot be accounted for by
the conventional coupled-channels calculation 
have been obtained \cite{NBD04,J02,D07,S08,piasecki}. 
One of these examples is the quasi-elastic scattering experiment for the
$^{20}{\rm Ne} + ^{90,92}{\rm Zr}$ systems \cite{piasecki}.
The experimental data show that 
the quasi-elastic barrier distribution 
for these systems behaves in a significantly different way from each other: 
the barrier distribution for the $^{20}$Ne+$^{92}{\rm Zr}$ system 
is much more smeared than that for the $^{20}$Ne+$^{90}{\rm Zr}$ 
system \cite{piasecki}. 
On the other hand, the coupled-channels 
calculations that take into account the collective rotational excitations in $^{20}$Ne 
as well as the vibrational excitations in $^{90,92}$Zr lead to similar barrier
distributions 
for both systems, 
because the strongly deformed 
$^{20}{\rm Ne}$ nucleus mainly
determines the barrier structure while the difference
in the collective excitations in the two Zr targets plays a minor role. 
In Ref. \cite{piasecki}, it was suggested that single-particle 
excitations in the colliding nuclei 
are responsible for smearing the barrier 
distribution for the $^{20}$Ne + $^{92}$Zr system. Notice that  
the single-particle excitations are expected to be much more
important for the $^{92}{\rm Zr}$ nucleus compared to the 
$^{90}{\rm Zr}$ nucleus, which has the $N = 50$ 
shell closure. 
In fact, while there are only 12 states in the $^{90}$Zr nucleus 
up to 4 MeV, there are 53 known states 
in the $^{92}$Zr nucleus \cite{bnl}. 
For 5 MeV, the number of known states is 35 and 87 for 
$^{90}$Zr and $^{92}$Zr, respectively (for another nucleus, $^{116}$Sn, 
there are 81 known levels up to 3.9 MeV and 112 levels up to 4.3 MeV 
\cite{RWK91,IWRK93}). 

The aim of this paper is to investigate 
the effect of low-lying 
single-particle excitations on low-energy heavy-ion reactions, as conjectured in Ref. \cite{piasecki}. 
In order to understand qualitatively the effect of non-collective 
excitations, in this paper 
we shall use a schematic model, that is, one-dimensional 
barrier penetration in the presence of the couplings to 
intrinsic degrees of freedom. 
The single-particle degrees of freedom can be 
described in several ways \cite{diaz,akw1,akw2,akw3,akw4,zagrebaev}. 
For instance, 
Ref.\cite{diaz} used the Lindblad approach to discuss 
the role of quantum decoherence in deep subbarier hinderance 
of fusion cross sections. 
In this paper, we employ the random matrix theory (RMT) 
to describe
the single-particle excitations (see Refs. \cite{PW07,rmtrev,rmtrev2,B97} 
for recent reviews on RMT).
The random matrix approach for heavy-ion reactions 
has been developed in 1970's by 
Weidenm\"{u}ller and his collaborators 
in order to analyse 
heavy-ion deep inelastic collisions (DIC)
\cite{akw1,akw2,akw3,akw4}.
At that time, they derived the transport coefficients based on RMT \cite{BNW79} 
and solved the classical transport equations (see also 
Refs. \cite{TNOY81,NT83}). 
The RMT has also been employed to discuss quantum dissipation 
\cite{W90,BDK96,MA97}. 
In this paper, instead of solving the classical equations, we 
directly solve the coupled-channels equations quantum mechanically 
by including the single-particle excitations described by RMT. 
Our approach is therefore similar to that in Ref. \cite{zagrebaev}, in 
which the coupled-channels equations with 200 dimension were solved 
for a one-dimensional model 
using a semi-classical approximation. 
In contrast to Ref. \cite{zagrebaev}, we apply our formalism to 
the subbarrier regime without using the semi-classical approximation. 
This will enable us to assess the effect of single-particle excitations 
on quantum tunneling and thus on the barrier distribution. 
By treating the single-particle states explicitly, we can also 
discuss the excitation spectra as a function of incident 
energy. 

The paper is organized as follows.
In Sec. II, we detail the coupled-channels formalism 
with single-particle excitations described by RMT. 
In Sec. III, we 
apply the formalism to one-dimensional models for quantum tunneling. 
We discuss the effect of single-particle excitations on 
the barrier penetrability, the barrier distribution, and 
the excitation spectra. 
Using the results for the one-dimensional model, we also 
discuss the effect of non-collective excitations on 
quasi-elastic barrier distribution for the $^{20}$Ne+$^{92}$Zr 
system. 
We then summarize the paper in Sec. IV.

\section {formalism}
\subsection{Coupled-channels method}

The aim of this paper is to discuss the effect of 
single-particle excitations on one-dimensional barrier 
penetrability. For this purpose, 
we assume the following Hamiltonian:
\begin{eqnarray}
  H = -\frac{\hbar^2}{2\mu}\frac{d^2}{dx^2}+ V_{\rm rel}(x) 
      + H_0(\xi) + V_{\rm coup}(x,\xi).
\end{eqnarray}
Here, 
$\mu$ is the reduced mass and $V_{\rm rel}(x)$ is a potential for the 
relative motion. 
$H_0(\xi)$ is a Hamiltonian for the intrinsic degrees of freedom 
of the colliding nuclei, 
and the last term, $V_{\rm coup}(x,\xi)$, is a coupling Hamiltonian 
between the relative motion and the internal degrees of freedom. 

The coupled-channels equations for this Hamiltonian are 
obtained by expanding
the total wave function in terms of the eigen functions of $H_0(\xi)$ and 
read,
\begin{eqnarray}
  \left\{-\frac{\hbar^2}{2\mu}\frac{d^2}{dx^2} + V_{\rm rel}(x)
  + \epsilon_n - E \right\}\psi_n(x)&&  \nonumber\\
  + \sum_m V_{nm}(x)\psi_m(x) &=& 0.
  \label{cceq}
\end{eqnarray}
Here, $\epsilon_n$ and $\psi_n(x)$ are the excitation energy and
the wave function for the $n$-th channel, respectively.
$V_{nm}(x)$ is a coupling matrix and is a function of
the coordinate $x$.

The coupled-channels equations are solved by imposing 
the boundary conditions of 
\begin{align}
  \psi_n(x) &\rightarrow \delta_{n,0}\,e^{-ik_0x} + r_n \,e^{ik_nx}\ \ \  
              {\rm for}\ \ \ x\rightarrow +\infty \label{bc1}\\
            &\rightarrow t_n \,e^{-ik_nx}\ \ \ {\rm for}\ \ \ 
             x\rightarrow -\infty, \label{bc2}
\end{align} 
where $\displaystyle k_n=\sqrt{{2\mu (E-\epsilon_n)}/{\hbar^2}}$ is 
the wave number for the $n$-th channel, and 0 represents the entrance channel.
We have assumed that 
the projectile is incident from the right hand side of the 
potential barrier. 
With the transmission coefficients $t_n$, 
the penetration probability for the inclusive process is calculated as 
\begin{equation}
P(E) = \sum_n P_n(E) = \sum_n\frac{k_n}{k_0} |t_n|^2. 
\end{equation}
The barrier distribution is obtained by taking the derivative of 
$P(E)$, that is, $dP(E)/dE$ \cite{HTB97}. 

In order to take into account the single-particle excitations, as we will show 
in the next section, one has 
to include a large number of channels. 
Since it is 
time and memory consuming to solve the coupled-channels equations 
with a large dimensionality, 
in this paper 
we employ a constant coupling
approximation \cite{dasso2}.
In this approximation, the coupling matrix 
is assumed to be a constant over the whole range of $x$.
Then, one can diagonalize the matrix $A=(V_{nm}+\epsilon_n\delta_{n,m})$
with a coordinate independent unitary matrix $U$,  
\begin{eqnarray}
  UAU^{\dagger} = diag\left\{\lambda_1,\lambda_2,\cdots\right\}
\end{eqnarray}
where, $\lambda_1,\lambda_2,\cdots$ are the eigenvalues of $A$.
Transforming the channel wave functions as  
\begin{eqnarray}
  \widetilde{\psi}_n(x) = \sum_m U_{nm}\psi_m(x), 
\end{eqnarray}
the coupled-channels equations are 
transformed to a set of the uncoupled equations,
\begin{eqnarray}
  \left\{-\frac{\hbar^2}{2\mu}\frac{d^2}{dx^2} + V_{\rm rel}(x)
    + \lambda_n - E\right\}\widetilde{\psi}_n(x) = 0.
\end{eqnarray}
We call the transformed channels the eigen-channels and, 
for each eigen-channel,
$V_n(x) = V_{\rm rel}(x) + \lambda_n$ the eigen-potential.

The boundary conditions 
for $\widetilde{\psi}_n(x)$ 
are given by 
\begin{eqnarray}
 \widetilde{\psi}_n(x) \rightarrow 
    U_{n0}\left(e^{-ik_0x}+\widetilde{r}_n e^{ik_0x}\right)\ \ \ 
                      {\rm for}\ \ \ x\rightarrow +\infty
    \label{bc4}
\end{eqnarray}
and
\begin{eqnarray}
  \widetilde{\psi}_n(x) \rightarrow U_{n0}\ \widetilde{t}_n\ 
                        e^{-ik_0x}\ \ \ {\rm for}\ \ \ 
                        x \rightarrow -\infty
    \label{bc5}
\end{eqnarray}
where the reflection and transmission coefficients are related to
the original coefficients in Eqs. (\ref{bc1}) and (\ref{bc2}) by
$\displaystyle \widetilde{r}_n=\sum_m {U_{nm}}r_m/{U_{n0}}$
 and 
$\displaystyle \widetilde{t}_n=\sum_m{U_{nm}}t_m/{U_{n0}}$, 
respectively.
Here, we have assumed that the excitation
energies are small compared to the incident energy so that 
$k_n$ can be approximated by $k_0$ \cite{dasso2}.
Using the coefficients $\widetilde{t}_n$,
the penetrability is calculated as 
\begin{eqnarray}
  P(E) = \sum_n |U_{n0}|^2|\widetilde{t}_n(E)|^2
  \label{cc_penet}.
\end{eqnarray}
The reflection coefficients in the original basis 
are given by 
\begin{equation}
r_n=\sum_m (U^{-1})_{nm}\,U_{m0}\,
\widetilde{r}_m,
\end{equation}
from which the $Q$-value distribution (that is, the 
excitation spectrum) is computed as
\begin{equation}
  f(\epsilon)=
  \sum_n
  \frac{k_n}{k_0} |r_n|^2\,\delta(\epsilon-\epsilon_n)
  \sim 
  \sum_n
  |r_n|^2\,\delta(\epsilon-\epsilon_n).
  \label{qdist}
\end{equation}

\subsection{Coupling matrix elements}

We solve the coupled-channels equations, (2), using the 
constant coupling approximation by including both collective and 
non-collective excitations. 
For the collective excitations, 
we assume either the vibrational or the rotational couplings. 
The coupling matrix for the vibrational 
coupling is given by 
\begin{eqnarray}
  (V_{nm}) = F
        \left(
          \begin{array}{rr}
            0  &  1         \\
            1  &  0 
          \end{array}
        \right),
        \label{Vcp_vib}
\end{eqnarray}
if we truncate the phonon space up to 1-phonon state \cite{ccfull}. 
Here, $F$ is a coupling constant, and we have assumed the linear coupling. 
For the rotational coupling, the coupling matrix 
is given by
\begin{eqnarray}
  \displaystyle
  (V_{nm}) &=& \frac{F_2}{\sqrt{4\pi}}
        \left(
          \begin{array}{ccc}
            0  &  1 & 0        \\
            1  &  \frac{2\sqrt{5}}{7} & \frac{6}{7} \\
            0  &  \frac{6}{7}  & \frac{20\sqrt{5}}{77}
          \end{array}
        \right) \nonumber \\
       &+& \frac{F_4}{\sqrt{4\pi}}
        \left(
          \begin{array}{ccc}
            0  &  0 & 1        \\
            0  &  \frac{6}{7} & \frac{20\sqrt{5}}{77} \\
            1  &  \frac{20\sqrt{5}}{77}  & \frac{486}{1001}
          \end{array}
        \right),
        \label{Vcp_rot}
\end{eqnarray}
up to the $4^{+}$ state in the rotational band\cite{ccfull}, where $F_2$ and $F_4$ are 
the quadrupole and hexadecapole coupling strengths, respectively. 

For the single-particle excitations, 
we consider an ensemble of coupling matrix elements 
based on the random matrix theory 
\cite{akw1,akw2,akw3}. 
We assume that the matrix elements are uncorrelated random
numbers obeying a Gaussian distribution with zero mean. 
That is, 
we require that the first and the second moments of the
coupling matrix elements satisfy the following equations
\cite{PLB62-248}
\begin{align}
  &\overline{V_{nm}(x)} = 0 \label{1stmom} \\
  &\overline{V_{rs}V_{nm}} = (\delta_{r,n}\delta_{s,m} +
                                    \delta_{r,m}\delta_{s,n})
                                   g_{nm}
   \label{2ndmom} \\
&  g_{nm}  = \frac{w_0}{\sqrt{\rho(\epsilon_n)\rho(\epsilon_m)}}
                 e^{-\frac{(\epsilon_n - \epsilon_m)^2}{2\Delta^2}},
    \label{fmfac2}
\end{align}
where the overbar denotes an ensemble average 
and $\rho(\epsilon)$ is the nuclear level density. 
Here, we have assumed the coordinate independent matrix elements 
according to the constant coupling approximation. 

For the single-particle excitations, we generate
the coupling matrix elements according to these equations many times.  
For each coupling matrix, 
we do not vary 
the matrix elements for the 
collective excitations, 
which 
are uniquely determined once 
the coupling is specified.
For each coupling matrix, we solve the coupled-channels 
equations and calculate the penetrability and the reflection 
probability. 
The physical results are then obtained by taking an average of 
these quantities. 

In the actual calculations shown in the next section, 
we discretize the quasi-continuum single-particle spectrum 
in the coupled-channels equations \cite{nemes,lipperheide} (see 
also Ref.\cite{KK74}).
Introducing the level density $\rho(\epsilon)$, 

\begin{eqnarray}
  \rho(\epsilon) = \sum_n \delta(\epsilon - \epsilon_n),
  \label{exact_ld}
\end{eqnarray}
the coupled-channels equation can be written in the following form :
\begin{eqnarray}
  \left\{-\frac{\hbar^2}{2\mu}\frac{d^2}{dx^2} + V_{\rm rel}(x)
  + \epsilon_n - E \right\}\psi_n(x)&&  \nonumber\\
  + \int d\epsilon \rho(\epsilon)V_{n\epsilon}(x)\psi_{\epsilon}(x) = 0.
  \label{cceq2}
\end{eqnarray}
In this equation, we 
assume a quasi-continuum spectrum for 
the single-particle 
excited states, and 
discretize the integral with a constant energy spacing, $\Delta\epsilon$. 
For the ground state and the collective excitation channels, 
we then obtain,
\begin{eqnarray}
  &&\left\{-\frac{\hbar^2}{2\mu}\frac{d^2}{dx^2} + V_{\rm rel}(x)
  + \epsilon_n - E \right\}\psi_n(x) \nonumber \\
  &&+ \sum_{m \notin {\rm sp}} V_{nm}(x)\psi_{m}(x)  \nonumber \\
  &&+ \sum_{m \in {\rm sp}} \Delta\epsilon
      \rho(\epsilon_m)V_{n\epsilon_m}(x)\psi_{\epsilon_m}(x) = 0,
  \label{cceq3}
\end{eqnarray}
while for the 
single-particle channels denoted by $\epsilon_n$ we obtain 
\begin{eqnarray}
  &&\left\{-\frac{\hbar^2}{2\mu}\frac{d^2}{dx^2} + V_{\rm rel}(x)
  + \epsilon_n - E \right\}\psi_{\epsilon_n}(x) \nonumber \\
  &&+ \sum_{m \notin {\rm sp}} V_{\epsilon_nm}(x)\psi_{m}(x)  \nonumber \\
  &&+ \sum_{m \in {\rm sp}} \Delta\epsilon
      \rho(\epsilon_m)V_{\epsilon_n\epsilon_m}(x)\psi_{\epsilon_m}(x) = 0.
  \label{cceq4}
\end{eqnarray}
Here, $m\notin$ sp denotes a summation over the ground state and the 
collective channels, while $m \in$ sp a summation
over the single-particle channels.
These equations can be expressed in a simpler way by 
multiplying a factor $\sqrt{\rho(\epsilon_n)\Delta\epsilon}$ for 
each index $\epsilon_n$ representing the single-particle channels.
That is, 
\begin{align}
  \widetilde{\psi}_{\epsilon_n}(x) 
             &= \sqrt{\rho(\epsilon_n)\Delta\epsilon}
               \ \psi_{\epsilon_n}(x), \\
  \widetilde{V}_{n\epsilon_m}(x) 
             &= \sqrt{\rho(\epsilon_m)\Delta\epsilon}
               \ V_{n\epsilon_m}(x), \\
  \widetilde{V}_{\epsilon_n\epsilon_m}(x) 
             &= \Delta\epsilon
               \sqrt{\rho(\epsilon_n)\rho(\epsilon_m)}
               \ V_{\epsilon_n\epsilon_m}(x). 
\end{align}
With these wave functions and the coupling matrix elements, 
Eqs. (\ref{cceq3}) and (\ref{cceq4}) read,
\begin{align}
&\left\{-\frac{\hbar^2}{2\mu}\frac{d^2}{dx^2} + V_{\rm rel}(x)
  + \epsilon_n - E \right\}\psi_n(x)  \nonumber \\
  &+ \sum_{m \notin {\rm sp}} V_{nm}\psi_m(x) 
   + \sum_{m \in {\rm sp}} 
      \widetilde{V}_{n\epsilon_m}(x)
      \widetilde{\psi}_{\epsilon_m}(x)
  = 0,  \label{cceq5}
\end{align}
and 
\begin{align}
  &\left\{-\frac{\hbar^2}{2\mu}\frac{d^2}{dx^2} + V_{\rm rel}(x)
  + \epsilon_n - E \right\}\widetilde{\psi}_{\epsilon_n}(x)  \nonumber \\
  &+ \sum_{m \notin {\rm sp}} \widetilde{V}_{\epsilon_nm}\psi_m(x) 
   + \sum_{m \in {\rm sp}}
     \widetilde{V}_{\epsilon_n\epsilon_m}(x)
     \widetilde{\psi}_{\epsilon_m}(x)
  = 0, \label{cceq6}
\end{align}
respectively. 
From Eqs. (\ref{bc1}) and (\ref{bc2}), the boundary conditions for 
$\widetilde{\psi}_{\epsilon_n}(x)$ are given by 
\begin{align}
  \widetilde{\psi}_{\epsilon_n}(x) &\rightarrow 
\sqrt{\rho(\epsilon_n)\Delta\epsilon}
                              \ t_n e^{-ik_n x} \nonumber \\
                         &= \widetilde{t}_n e^{-ik_n x}
                            \ \ \ {\rm for}\ \ x \rightarrow -\infty
\end{align}
and
\begin{align}
  \widetilde{\psi}_{\epsilon_n}(x)  &\rightarrow \sqrt{\rho(\epsilon_n)\Delta\epsilon}
                              \ r_n e^{ik_n x}\nonumber \\
                        &= \widetilde{r}_n e^{ ik_n x}
                            \ \ \ \ {\rm for}\ \ x \rightarrow +\infty,
\end{align}
where, 
$\widetilde{t}_n = \sqrt{\rho(\epsilon_n)\Delta\epsilon}\ t_n$ and 
$\widetilde{r}_n = \sqrt{\rho(\epsilon_n)\Delta\epsilon}\ r_n.$
The penetrability is then given by
\begin{align}
  P(E) &= 
        \sum_{n\notin {\rm sp}} \frac{k_n}{k_0}|t_n|^2
                  + \sum_{n\in {\rm sp}}\frac{k_n}{k_0}
                    \rho(\epsilon_n)\Delta\epsilon\ |t_n|^2, \nonumber\\
       &= \sum_{m\notin {\rm sp}}\frac{k_n}{k_0}|t_n|^2
                  + \sum_{n\in {\rm sp}}\frac{k_n}{k_0}|\widetilde{t}_n|^2.
\end{align}

This method with a constant energy step 
considerably reduces the computation time 
as compared to the case of treating the exponentially increasing 
number of single-particle levels as they are. 
It also validates the use
of the random matrix theory.
An important assumption in RMT is that the ensemble average of
a quantity is equivalent to the energy average of that quantity over the
spectrum \cite{rmtrev}.
In our case, we expect that the ensemble average of the 
calculated results 
corresponds to 
the energy average of the same quantities within the energy 
spacing of $\Delta\epsilon$.

\section{Results}

\subsection{Vibrational coupling}

\begin{figure}[hb]
  \begin{center}
    \includegraphics[clip,width=50mm,height=60mm]{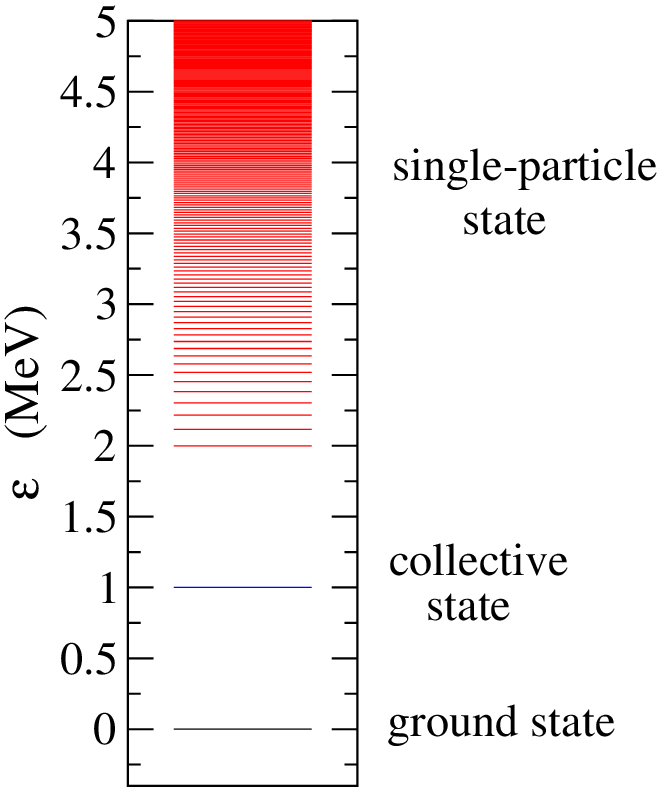}
    \caption{(Color online) The energy spectrum for the model 
calculation we employ. 
There is a collective vibrational state at 1 MeV, while 
single-particle states exist from 2 MeV with an exponentially 
increasing level density. }  
    \label{fig:spectrum}
  \end{center}
\end{figure}

We now numerically solve the coupled-channels equations and 
discuss the effect of single-particle excitations on 
barrier penetrability. 
We first consider the vibrational coupling. 
We assume that there is a collective vibrational state at 
1 MeV, whose coupling to the ground state is 
given by Eq. (\ref{Vcp_vib}) with 
$F = 2$ MeV. 
For the single-particle states, 
we consider 
a level density given by 
$\rho(\epsilon) = \rho_0\ e^{2\sqrt{a\epsilon}}$ with  
$\rho_0 = 0.039\ {\rm MeV^{-1}}$ and $a = 29/8\ {\rm MeV^{-1}}$, 
starting from 2 MeV. 
The value of $\rho_0$ was determined so that the number of 
single-particle levels 
is 200 up to 5 MeV. 
For the 
parameters for the couplings in Eq. (\ref{fmfac2}), we 
follow Ref. \cite{PLB62-248} to use $\Delta$=7 MeV. 
We arbitrarily choose the coupling strength to be 
$w_0 = 0.005$ MeV. 
The energy spectrum for this model is shown in Fig.\ref{fig:spectrum}.
For the potential for the relative motion, $V_{\rm rel}(x)$, 
we use a Gaussian function
\begin{equation}
  V_{\rm rel}(x) = V_{\rm B} e^{-\frac{x^2}{2s_0^2}}, 
\label{relgaus}
\end{equation}
with $V_{\rm B}$ = 100 MeV and $s_0$ = 3 fm \cite{dasso2}. 
The reduced mass $\mu$ is taken to be 29$m_N$, $m_N$ being the 
nucleon mass. 

\begin{figure}[t]
  \begin{center}
    \begin{minipage}[t]{75mm}
      \includegraphics[clip,width=80mm,height=60mm]{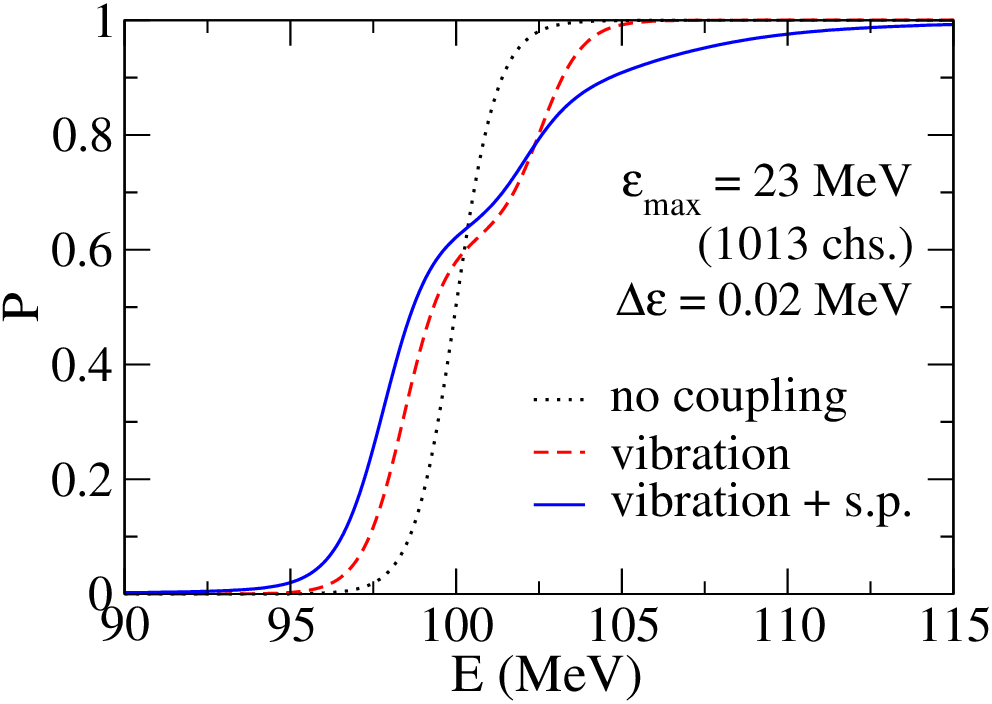}
      \caption{(Color online) The potential penetrability obtained with several methods. 
The dotted line is obtained without channel coupling, while 
the dashed line takes into account only the collective vibrational 
excitation. The solid line shows the result with 
both the collective and the single-particle excitations.}
      \label{fig:penet}
    \end{minipage}
    \hspace{0.5cm}
    \begin{minipage}[t]{75mm}
       \includegraphics[clip,width=80mm,height=60mm]{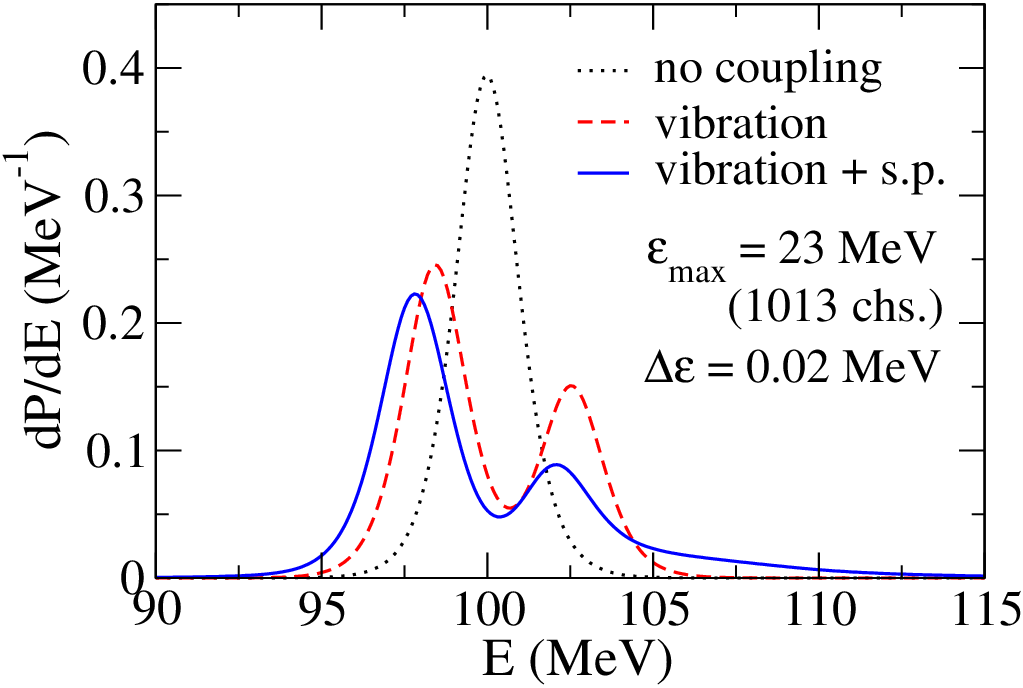}
       \caption{(Color online) The barrier distribution defined by the first derivative 
of the penetrability. The meaning of each line is the 
same as in Fig.\ref{fig:penet}.}
       \label{fig:bardist}
    \end{minipage}
  \end{center}
\end{figure}

Figure \ref{fig:penet} shows the penetrabilities thus obtained.
The corresponding barrier distributions 
are shown in 
Fig. \ref{fig:bardist}. 
The dotted and the dashed lines show the results  
without the channel couplings and those only with the collective 
excitation, respectively. 
The solid line shows the results with both the single-particle excitations and 
the collective excitation. 
We include the single-particle states up to $\epsilon_{\rm max}=$23 MeV with 
energy spacing of $\Delta \epsilon$=0.02 MeV. 
With this model space, the number of channels included is 1013 (we 
treat the low-lying single-particle states as discrete states when the energy spacing 
is larger than $\Delta \epsilon$). 
This result is obtained by 
generating 
the coupling matrix elements 30 times to take an ensemble average. 
We have found that the fluctuation around the average is small. For instance, 
at $E=100$ MeV, the averaged penetrability is $P=0.622$, whereas 
the root-mean-square deviation is 4.366$\times 10^{-3}$. 
As we mentioned in the previous section, the results shown in Fig. 2 
are obtained 
with the constant coupling approximation. 
With a smaller value of 
$\epsilon_{\rm max}$, we have solved the coupled-channels equations 
exactly and have confirmed that the constant coupling 
approximation works qualitatively well. 

The collective excitation leads to a double peaked structure 
of barrier distribution. 
One can see that the single-particle excitations 
suppress the penetrability at energies above the barrier, and 
at the same time smear the higher energy peak in the 
barrier distribution, although the main structure of the barrier 
distribution is still determined by the collective excitation. 
The single-particle excitations also lower the barrier and thus 
increase the penetrability at energies below the barrier, due to 
the well-known potential renormalization \cite{THAB94}. 

\begin{figure}[t]
    \includegraphics[clip,width=50mm,height=100mm]{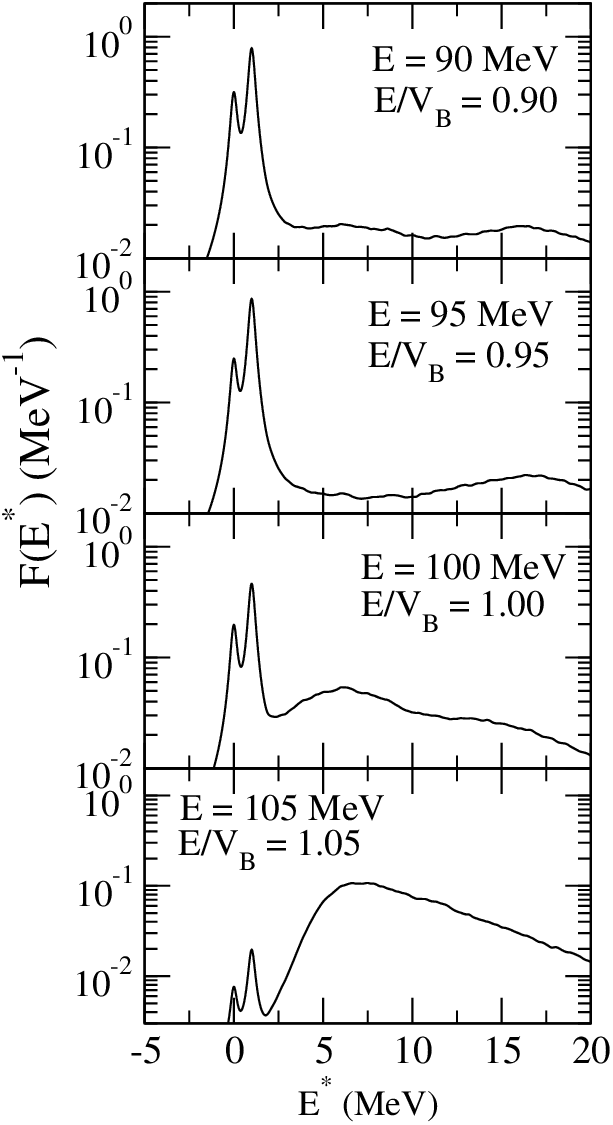}
    \caption{(Color online) The $Q$-value distribution for the reflected flux at four energies 
as indicated in the figure. It is obtained by smearing the discrete 
distribution with a Lorentzian function with the width of 0.2 MeV. 
The peaks at $E^{*}$=0 MeV and
      1 MeV correspond to the elastic and the 
collective excitation channels, respectively.}
    \label{fig:qdist}
\end{figure}

The $Q$-value distribution for the reflected flux is shown in 
Fig. \ref{fig:qdist} at four incident energies indicated in the figure. 
For a presentation purpose, we fold the discrete 
distribution with a Lorentz function, 
\begin{eqnarray}
  g(\epsilon) = \frac{1}{\pi}\frac{\eta}{\epsilon^2 + \eta^2},
\end{eqnarray}
with the width of $\eta=$0.2 MeV. 
That is, with the function defined by Eq. (\ref{qdist}), we compute 
\begin{align}
  F(E^{*}) &= \int d\epsilon f(\epsilon) g(E^{*}-\epsilon)  \nonumber \\
           &= \sum_n |r_n|^2
              \frac{1}{\pi}\frac{\eta}{(E^{*}- \epsilon_n)^2 + \eta^2}.
\end{align}
In the figure, 
the peaks at $E^{*}=0\ {\rm MeV}$ and $E^{*}=1\ {\rm MeV}$ correspond to the 
elastic channel and the collective excitation channel, respectively.
One can see that at energies well below the barrier 
the elastic and the collective peaks dominate in the distribution. 
As the energy increases, the 
single-particle excitations become more and more important. 
This behaviour is consistent with the experimental $Q$-value 
distribution observed for $^{16}$O+$^{208}$Pb \cite{evers,lin} 
and $^{16}$O+$^{184}$W \cite{lin} reactions.  
At energies above the barrier, the single-particle contribution is even larger than 
the contribution of the elastic and the collective peaks. 

\subsection{Rotational coupling}

Let us next consider the rotational coupling.
For this purpose, we mock up the $^{20}$Ne + $^{92}$Zr system. 
That is, we consider the rotational excitations in $^{20}$Ne up to the 4$^+$ 
state and the single-particle excitations in $^{92}$Zr. 
The energies of the rotational states are thus 
$\epsilon_{2^{+}}$ = 1.634 MeV and $\epsilon_{4^{+}}$ = 4.248 MeV for the 
2$^+$ and 4$^+$ states, respectively. 
The values of the coupling strengths $F_2$ and $F_4$ in Eq. (\ref{Vcp_rot})
are estimated with the collective model for the coupling form factor 
at the barrier position with the deformation 
parameters of $\beta_2$ = 0.46 and $\beta_4$ = 0.27. 
This yields $F_2=-6.892$ MeV and $F_4=-4.632$ MeV. 
For the single-particle excitations in $^{92}$Zr, we 
consider the energy range of 2 MeV $\leq \epsilon \leq$ 16 MeV, with 
the exponential level density with 
$\rho_0 = 0.039$ MeV$^{-1}$ and $a = 30/8$ MeV$^{-1}$.
For the coupling strength, we use 
$w_0$ = 0.005 MeV and $\Delta = 4.0$ MeV. 
These parameters are adjusted so that the rotational excitation 
in $^{20}$Ne gives the main structure of the barrier distribution. 

In the calculation shown below, we also include the mutual 
excitations of the projectile and the 
target nuclei. 
In order to avoid closed channels, we introduce the energy cutoff 
and include those channels whose total excitation energy is below 16 MeV. 
With this set-up, the number of channels included is 1688. 

We use the Gaussian function for the potential for the relative motion 
with $V_B$ = 51.76 MeV and $s_0$ = 2.475 fm.
This yields the same barrier height and the curvature as 
those with a Woods-Saxon potential 
with the parameters of $V_0$ = 59.9 MeV, $r_0$ = 1.2 fm, and 
$a$ = 0.63 fm for the $^{20}$Ne + $^{92}$Zr system. 
The reduced mass is taken to be 20$\times$ 92 $m_N$/112. 

\begin{figure}[t]
  \begin{center}
    \begin{minipage}[t]{75mm}
      \includegraphics[clip,width=80mm,height=60mm]{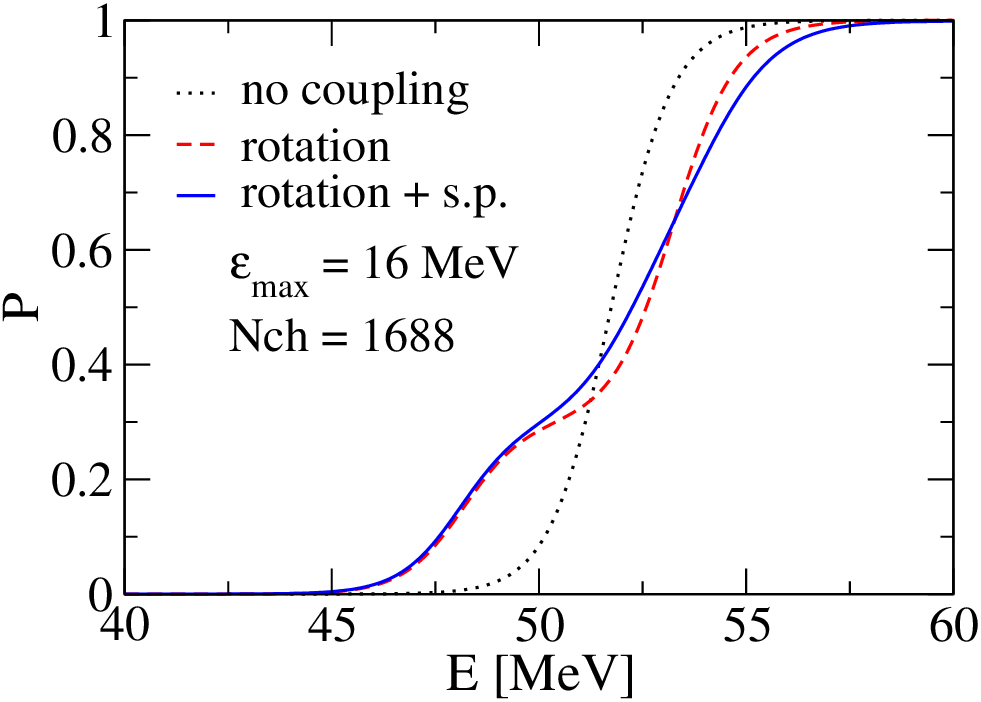}
      \caption{(Color online) 
The same as Fig. \ref{fig:penet}, but for the rotational coupling. }
      \label{fig:penet_rot}
    \end{minipage}
    \hspace{0.5cm}
    \begin{minipage}[t]{75mm}
       \includegraphics[clip,width=80mm,height=60mm]{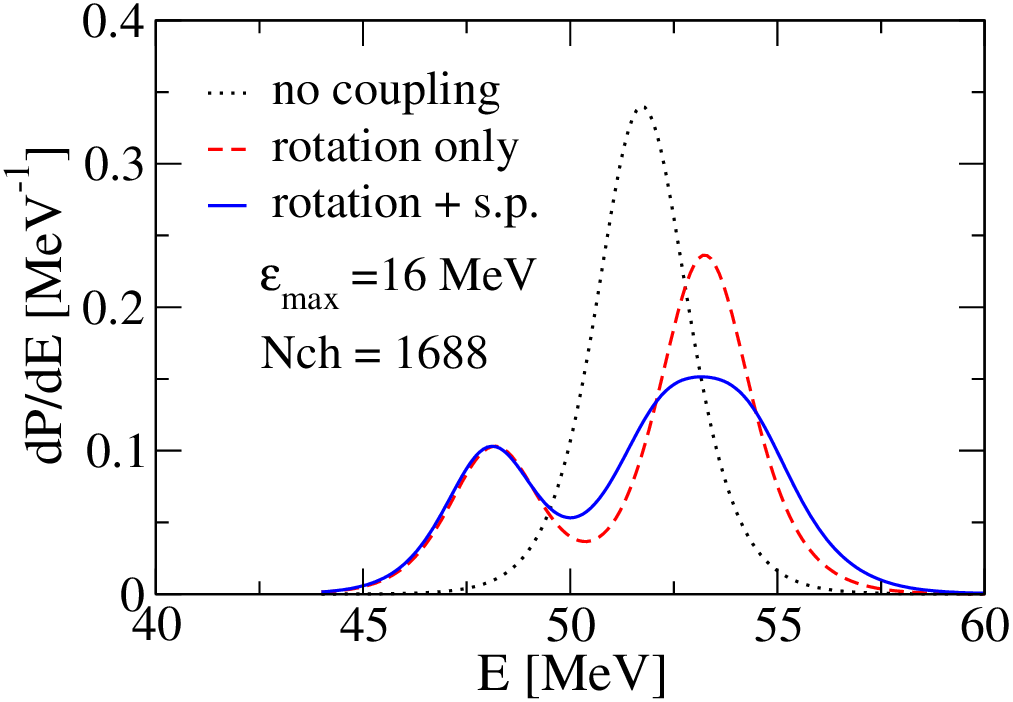}
       \caption{(Color online) The same as Fig. \ref{fig:bardist}, but for the rotational coupling. }
       \label{fig:bardist_rot}
    \end{minipage}
  \end{center}
\end{figure}

Figs. \ref{fig:penet_rot} and \ref{fig:bardist_rot} show the 
penetrability and the barrier distribution for this model, respectively. 
The meaning of each line is the same as in Figs. \ref{fig:penet} and
\ref{fig:bardist} for the vibrational coupling. 
With only the collective rotational excitations, 
there are three eigenbarriers whose height is 
48.2, 53.3, and 55.6 MeV. 
In contrast to the vibrational coupling, for the rotational coupling 
with a prolate deformation the main peak in the barrier distribution is 
not the lowest energy peak. 
For the parameters we use, the highest energy barrier carries a 
relatively small weight and the barrier distribution 
has only two visible peaks. 
The effect of single-particle excitations on the barrier distribution is 
similar to the case for the vibrational coupling and smears the 
higher energy peak in the barrier distribution.

\subsection{Quasi-elastic barrier distribution}

Using the eigenbarriers and the corresponding weight factors obtained 
in the previous subsection, one can compute the 
quasi-elastic scattering cross sections and the quasi-elastic 
barrier distribution in a three-dimensional space. 
That is, in the eigen-channel representation, the quasi-elastic 
scattering cross section is given by \cite{qelbar,ARN88},
\begin{equation}
\sigma_{\rm qel}(E,\theta) = \sum_i w_i \sigma_{\rm el}(E-\lambda_i,\theta),
\end{equation}
where $w_i=|U_{i0}|^2$ is the weight factor for the $i$-th 
eigen-channel and $\sigma_{\rm el}$ is the elastic scattering 
cross section. 
In order to calculate the elastic scattering cross sections, we 
use the Woods-Saxon potential indicated in the previous subsection. 
For the imaginary part of the optical potential, we assume an internal 
absorption, in which the imaginary part is well 
localized only inside the Coulomb barrier. 

\begin{figure}[t]
  \begin{center}
      \includegraphics[clip,width=80mm,height=60mm]{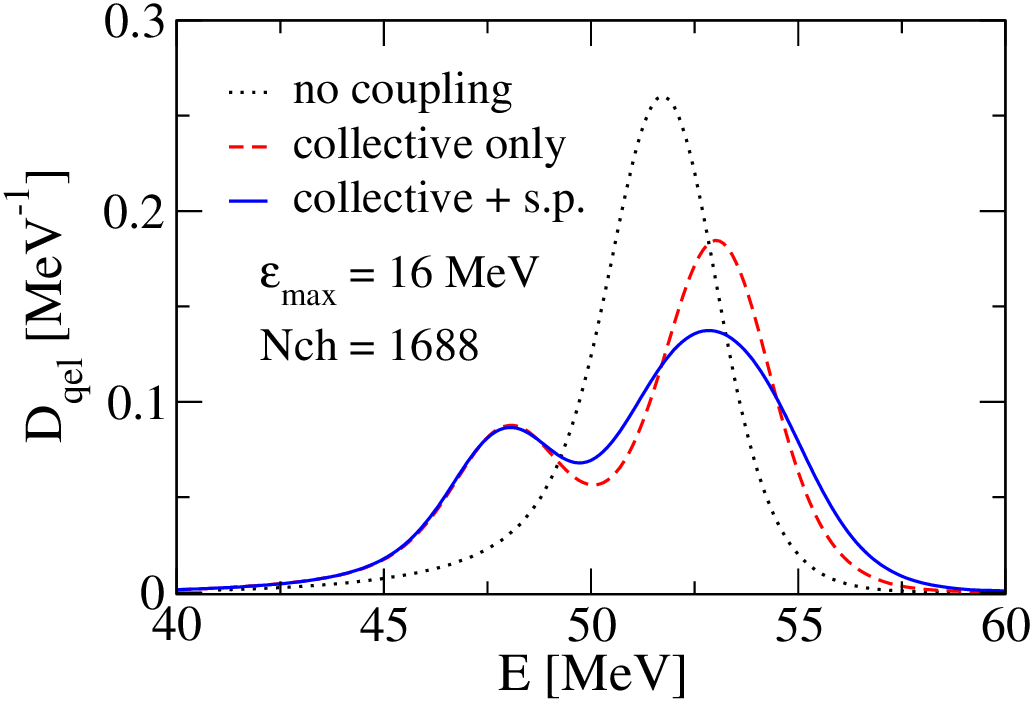}
      \caption{(Color online) The quasi-elastic barrier distribution 
for the $^{20}$Ne+$^{92}$Zr  system calculated with the 
weight factors and the eigenbarrier heights for the one-dimensional 
rotational model discussed in Sec. III B.  
The meaning of each line is the same as in Fig. \ref{fig:bardist}. }
      \label{fig:qel_bardist}
  \end{center}
\end{figure}

Fig. \ref{fig:qel_bardist} shows the quasi-elastic 
barrier distribution obtained with the same $w_i$ and $\lambda_i$ as in the previous subsection 
for the one-dimensional model. 
We use the point difference formula with $\Delta E_{\rm c.m.}$= 
2 MeV to calculate the quasi-elastic barrier distribution. 
The meaning of each line is the same as in Fig. \ref{fig:bardist}.
Because the quasi-elastic barrier distribution is by itself smeared 
more than the fusion barrier distribution \cite{qelbar}, and also 
because we use the point difference formula rather than taking 
the derivative, the higher energy 
peak in the barrier distribution is more smeared by the single-particle 
excitations as compared to the one-dimensional calculation shown in 
the previous subsection. 
The difference between the dashed line (the collective excitations only) 
and the solid line (the collective + single-particle excitations) 
is similar to the difference in the experimental 
quasi-elastic barrier distribution between 
$^{20}$Ne+$^{90}$Zr  and $^{20}$Ne+$^{92}$Zr  
systems. 
We therefore conclude that the single-particle excitations 
indeed provide a promising way to explain the difference in 
the quasi-elastic barrier distribution for the 
$^{20}$Ne+$^{90,92}$Zr  systems.

\section{Summary}

We have 
studied the role of 
single-particle excitations in 
heavy-ion reactions
at energies close to the Coulomb barrier. 
To this end, we 
employed a random matrix theory 
to describe the single-particle degrees of freedom. 
We applied the 
model to one dimensional barrier penetration problems by using 
the constant coupling approximation. 
In addition to the single-particle excitations, we also 
included the collective excitations with either a vibrational or 
a rotational character. 
We calculated the potential penetrability, the
barrier distribution and the $Q$-value
distribution.  
We also calculated the quasi-elastic barrier distribution 
for the $^{20}$Ne + $^{92}$Zr system 
using the eigenbarriers and their weight factors 
obtained with the one-dimensional model.
Our calculations show that
the single-particle excitations 
hinder the penetrability at energies above the barrier and
smear the high energy part of the barrier distribution.
In the $Q$-value distribution, we found that the
contribution from the single-particle 
excitations increases significantly as the 
incident energy increases. 

The experimental quasi-elastic 
barrier distributions are considerably different between 
$^{20}$Ne + $^{90,92}$Zr systems, despite that the coupled-channels 
calculations with collective excitations in the colliding 
nuclei lead to similar barrier distributions to each other. 
Our calculations imply that the difference can be indeed 
accounted for by the non-collective excitations in the target 
nuclei, as has been conjectured in Ref. \cite{piasecki}

In order to make a quantitative comparison to the experimental 
data and draw a definite conclusion on the quasi-elastic 
barrier distribution for the $^{20}$Ne + $^{90,92}$Zr systems, 
it will be an interesting future work to 
extend our study presented in this paper to three-dimensional 
calculations without resorting to the constant coupling 
approximation. 
This is so espeically because the constant coupling approximation 
which we employed in this paper may introduce a significant 
phase error and thus leads to an inconsistent angular dependent 
interference between different partial waves. 
It may also be 
interesting to 
see whether this model 
accounts for the hindrance of fusion cross sections at 
deep sub-barrier energies recently found in several systems. 
For this purpose also, we would have to 
take into consideration the coordinate dependence of the 
coupling form factor, especially around the touching point of the 
colliding nuclei\cite{ichikawa}. 

\begin{acknowledgments}
We thank E. Piasecki, M. Dasgupta, D.J. Hinde, 
M. Evers, and V.I. Zagrebaev for discussions. 
This work was supported
by the Grant-in-Aid for Scientific Research (C), Contract No.
22540262 from the Japan Society for the Promotion of Science.
\end{acknowledgments}

\end{document}